# Effect of Impurities on Pentacene Thin Film Growth for Field-Effect Transistors

*Elba Gomar-Nadal, Brad R. Conrad, William G. Cullen, Ellen D. Willams**

Physics Department and Materials Research and Science and Engineering Center,

2120 Physics Building, University of Maryland

College Park, MD 20742, USA

Email: edw@umd.edu

Fax: 001 301 314 9465

Phone: 001 301 405 6156

ABSTRACT

Pentacenequinone (PnQ) impurities have been introduced into a pentacene source material at number densities from 0.001 to 0.474 to quantify the relative effects of impurity content and grain boundary structure on transport in pentacene thin-film transistors. Atomic force microscopy (AFM) and electrical measurements of top-contact pentacene thin-film transistors have been employed to directly correlate initial structure and final film structures, with the device mobility as a function of added impurity content. The results reveal a factor four decrease in mobility without significant changes in film morphology for source PnQ number fractions below ~0.008. For these low concentrations, the impurity thus directly influences transport, either as homogeneously distributed defects or by concentration at the otherwise-unchanged grain boundaries. For larger impurity concentrations, the continuing strong decrease in mobility is correlated with decreasing grain size, indicating an impurity-induced increase in the nucleation of grains during early stages of film growth.

Introduction

During the last decade, dramatic advances have been made in the performance of organic thin film field-effect transistors (OTFTs), and their field-effect mobilities have exceeded those of transistors based on amorphous silicon[1, 2]. Despite the fast-paced progress, fundamental questions related to the mechanism and limiters of device operation remain unanswered. Both thin film morphology and chemical impurities have been identified as limiting charge carrier mobility. Studies with single crystals have shown the strong effect of small concentrations of impurities[3-6]. Studies of thin-film growth have revealed the mechanisms underlying grain formation in thin films as well as self-driven polycrystallization[7, 8]. However, questions still remain as to the specific roles these impurities play.

Reducing the concentration of quinones, which are the dominant impurity in acenes, has been demonstrated to correlate with improved mobility in single crystals[4, 6]. The quinone impurity content in commercial acenes has been reported to be about 0.7%[4], however there is also a significant enhancement of the quinone concentration in the near surface region of crystalline pentacene[6]. Thus for single crystals, the mechanisms by which impurities reduce mobility may involve both impurities located in the bulk of the crystal and at interface sites and structural defects such as dislocations[9-11]. Because of their increased size and non-planar structure, quinones may degrade mobility through the creation of structural defects. These defects in turn may affect film stability and moisture sensitivity, and induce local potentials with further effect on transport[12, 13].

In thin film transistors, the growth of the organic semiconductor results in a polycrystalline structure with typically much lower mobilities than that obtained for single-crystal devices[2]. The lower mobility has been attributed to the influence of grain boundaries and dislocations as sites for charge traps[14-17]. Structural defects in the thin film arise from the growth properties. For pentacene on $SiO_2$, growth occurs via nucleation with initial two-dimensional growth[18]. The critical nucleus size is small, on the order of two or three molecules, and the subsequent domain growth varies from compact to ramified depending on growth temperature and flux[19-21]. Growth of the second layer begins before completion of the first layer, and continuing growth is increasingly three-dimensional due to an Ehrlich-Schwoebel-type barrier that prevents diffusion across terrace edges[7]. This growth behavior is common and has been seen in many other types of systems[18, 22]. Since the majority of charge carriers in a thin film transistor are located at the semiconductor-dielectric interface[23, 24], a detailed knowledge of the morphology of the organic thin film at the interface with the dielectric is crucial to understand charge transport.

### Experimental Methods

The experiments were performed by introducing controlled amounts of one chosen type of chemical impurity – 6,13-pentacenequionene (PnQ) – in pentacene (Pn). PnQ (99%, Ref.246883) is an oxidative form of Pn used as starting material in the chemical reaction to produce Pn and is its main impurity[4]. The solid mixtures were prepared by mechanically mixing PnQ with commercial Pn under a dry nitrogen atmosphere in a glove box. The mixtures were made by consecutive solid dilution, starting with a 50% Pn/PnQ mixture, by adding commercial Pn. The mixture components were ground finely with a glass mortar and pestle with repeated grinding and mixing. The compositions tested covered a range of added PnQ from weight percentage +0.0 to +50%, equivalent to PnQ number fractions ranging from 0.0 to 0.474. The PnQ thin film composition will be different from the PnQ source material composition and decrease across the thin film thickness since PnQ has a slightly lower evaporation temperature than Pn[4]. However, the film composition will be directly proportional to the percentage of PnQ present in the source material for films deposited under the same experimental conditions.

A series of top-contact Pn OTFTs were prepared using the PnQ/Pn solid admixtures as the source material. Prime grade silicon wafers ($p^+$-Si) with 300nm (±3%) thermally grown oxide were used as device substrates. The $p^+$-Si/$SiO_2$ substrates were cleaned by sonication in acetone and isopropanol (IPA) for 5 minutes, rinsed with IPA and dried with nitrogen. Deposition was performed at 0.09 Å/s at $10^{-7}$ Torr pressure with the substrate at room temperature. The source materials were increased to deposition temperature (~195ºC) over a 15 min. interval. All of the films were prepared under the same deposition conditions. In addition, some OTFTs were prepared using Pn purified by heating it at a temperature slightly lower than its sublimation temperature for one hour *prior* to the thin film deposition. This method reduces impurities, such as PnQ, that have a sublimation temperature lower than pentacene. The results for these samples are noted as "cleaned Pn" in the figures.

The electrical characterization was performed on films with an equivalent of 50 nm of material deposition, under a nitrogen atmosphere. For OTFT device fabrication, top-contact electrodes (100 nm) were deposited by evaporating gold ($<10^{-6}$ Torr) through a shadow mask with channel length L = 100 μm and width W = 3000 μm. The reported field-effect mobilities μ are the average of at least 4 transistors for added PnQ > +0.5% and the average of at least 8 transistors for added PnQ ≤ + 0.5%. Analysis of the transport data to extract the device parameters followed standard procedures[25]. The mobilities reported are based on the linear mobilities measured for gate voltages between -60 V to -40 V. Mobility values are reported as normalized with respect to the measured mobility of the commercial pentacene (0.0% *added* PnQ) which is 0.11 ± 0.02 cm$^2$/Vs. Error bars are reported as the standard deviation (one sigma) of the repeated measurements. Tapping mode AFM was conducted using a DI MultiMode with silicon cantilevers on the device conduction channels after electrical measurement. Additional submonolayer films were also grown and imaged to facilitate nucleation and grain size analysis. Additional phase images were recorded for many samples.

**Results/ Discussion**

Transfer curves and normalized linear mobility as a function of the number fraction of added PnQ present in the source material for these transistors are shown in Fig. 1. The threshold voltages and ON/OFF ratios were comparable for transistors fabricated from all the different material compositions, with a $V_t$ average value of -12 ± 2 V and On/Off ratios of $10^4$ (see supplemental Table S1). Beyond an added number density of 0.007 PnQ, the mobility rapidly decreases for low concentrations of PnQ and saturates to low mobility values at high PnQ concentrations as can be seen in Fig. 1b. The inset of Fig. 1b focuses on the rapid degradation of the mobility with small amounts of added PnQ. The PnQ/Pn admixture device data is fit to a linear function and plotted in the inset of Fig. 1b, and can be extrapolated to zero mobility at a number fraction of 0.013 ± 0.004. Also from the inset, we can extrapolate to the reported[4] PnQ content of commercial Pn and see that our data for cleaned and commercial Pn is in good agreement with the expected value of 0.7 ± 0.1% PnQ impurity in the commercial material. All further PnQ concentrations will be quoted in absolute number fractions in the source, or added PnQ plus the native 0.7% PnQ by mass in commercial Pn. To assess the influence of PnQ impurity on the pentacene thin-film morphology, tapping mode AFM images were recorded for these films. The structures observed for a representative subset of impurity concentrations are shown in Fig. 2a. Samples prepared with purified Pn and with a PnQ number fraction of up to 0.008 present similar crystalline grain morphology, with grain sizes (~1μm) significantly larger than those with PnQ number fractions higher than 0.008. As the content of added impurity exceeds number fractions of 0.008, the samples show a drastic change in grain morphology, including a dramatic decrease in grain size. Together with the small grains, a low density of high-aspect-ratio protruding structures (appearing as white contrast in the AFM images in Fig 2a, and shown in 3-d in Fig 2b) becomes apparent around a PnQ number fraction of 0.041. These structures are similar in density and shape to those that occur when equivalent amounts of pure PnQ are deposited on clean $SiO_2$[26], suggesting phase separation of higher concentrations of PnQ from Pn[27, 28].

The correlation between the average grain size and the mobility for all the samples measured is shown in Fig. 2c. The grain sizes are found by using image processing to outline the irregularly shaped grains, finding the areas, and taking the average. The reported grain size is the diameter of an assumed circular grain with the average measured grain size. It is notable that the first rapid decrease in mobility by about a factor of four occurs with no significant change in grain size. Only when the amount of added impurity exceeds a PnQ number fraction of 0.008 does a correlated decrease in grain size also occur. To quantify the effects of the impurity on the growth process and changes in mobility, the thin film structures in the early stages of growth were also measured. In agreement with the literature, the early stages of growth of commercial Pn (a PnQ number fraction of 0.006) occur via layer-by-layer growth, and the first two monolayers are at least 90% completed before the next monolayers start to grow (Fig. SI3). The structures of films grown with a deposition time equal to that yielding 1 monolayer of Pn with a variable PnQ number fraction are shown in Fig. 3a. The morphology trends seen in the submonolayer thin film images, such as Fig. 3a, continue through the growth of the first few complete conduction channel layers, which are responsible for most of the charge transport[24]. Samples prepared with PnQ number fractions at or below 0.008 show the same behavior-formation of an almost complete first monolayer (approximately 15Å thick) and some nucleation sites of the second monolayer. Samples with PnQ number fractions higher than 0.008 display incomplete Pn coverage, and the formation of multi-layer crystallites with both elongated and rounded shapes (10 to 100 nm tall and 10 to few 100 nm long). AFM images of 100% PnQ films (with a deposition time equivalent to 1 Pn monolayer) yield the same type of crystallites with comparable shapes and dimensions covering 20% of the substrate. This confirms that there is substantial material segregation between PnQ and Pn during the film deposition.

As shown in Fig. 3b, at a 1 Pn monolayer deposition time, the fraction of the substrate area covered with PnQ initially increases with a slope slightly greater than one as a function of percentage of added PnQ in the source material, consistent with PnQ subliming more rapidly than Pn. At approximately 10% areal coverage, the rate of increase of the PnQ areal coverage decreases dramatically, and the areal coverage saturates at about 20%. This is the result of increasing multi-layer growth in the PnQ crystallites. Referring to Fig. 1b, the mobility drops by a factor of 10 over the PnQ composition range where the areal coverage of PnQ increases only to 10%. Thus the mobility decreases below 10% PnQ impurity cannot be due simply to PnQ-crystallite-induced loss of percolative pathways through the Pn regions.

Further information about the impact of the PnQ impurity on the pentacene film growth is provided by evaluating the nucleation density. This was accomplished by growing films until just before coalescence of adjacent growing grains begins and measuring the size and density of the grains. Nucleation density as a function of PnQ concentration in source material for a deposition time equivalent to 0.3-0.4 ML of Pn is summarized in Table 1. In agreement with the observations for bulk thin film morphology (Fig. 2a), purified samples and samples with PnQ number fractions between 0.006 (commercial Pn) and 0.008 show comparable island number (~ 3 islands/square micron, of area 0.13-0.14 square micron), as well as shape and spatial distribution[29]. Samples with a PnQ number fraction content above 0.008 have a significantly larger number of nucleation points and correspondingly smaller island areas.

The evolution of the growth morphology is consistent with a limited solid solubility of PnQ in Pn. If a PnQ number fraction up to about 0.008 can be incorporated in (or at the edges of) growing Pn islands with little disturbance in the long range crystal structure as suggested in Fig 4a and 4b, then the growing thin film morphology would be undisturbed up to that impurity number fraction. The rapid decrease in mobility with increasing impurity content in this range would then be due to direct increased scattering due to effects of the PnQ defects either within the Pn crystalline lattice or to increased concentration of PnQ at the island boundaries. The latter effect could hinder interconnections between adjacent islands, thus reducing favorable paths for e conduction[29]. The dramatic increase in the nucleation density of Pn domains above a PnQ number fraction of 0.008 could occur if critical nucleus formation were enhanced by impurity molecules, e.g. a critical nucleus might consist of Pn+PnQ. The increase in nucleation density would occur with increasing probability, as observed, with increasing PnQ content. Alternatively, the increased nucleation density could be explained as a PnQ-induced reduction in diffusion length for Pn. The mechanism by which this might occur is unclear, however it is known that Pn nucleation is highly sensitive to the surface composition[30-32]. Finally, increased nucleation could occur if PnQ decoration of growing Pn island edges inhibits incorporation of Pn. This would encourage formation of additional nucleation sites from the unincorporated Pn. The two likely mechanisms for impurity PnQ effects on Pn growth, therefore, will be differentiated by the distribution of PnQ in the film. In one case, PnQ would be distributed relatively uniformly throughout the Pn grains. In the other case, PnQ would be concentrated at the grain boundaries. Phase images revealed no additional information concerning the specific locations of the PnQ molecules. More information, however, can be extracted from detailed analysis of the growth, which will be presented elsewhere[33, 34].

In evaluating whether the primary cause of the PnQ induced mobility degradation is due to PnQ effect inside or on the perimeter of grain boundaries, it is useful to evaluate PnQ distributions that would be required for each case. From the inset of Fig. 1, a linear function can be fit and extrapolated to zero mobility at a number fraction of $0.013 \pm 0.004$. For the hypothesis that PnQ is uniformly distributed, this would indicate that a single PnQ molecule effects the charge transport of approximately 80 Pn molecules or a circular radius of ~5 Pn molecules, as illustrated in Fig 4a. For the alternative hypothesis, where PnQ is located primarily on the grain boundaries, the effect of PnQ will depend on the grain size. A Pn grain area of 0.250 $\mu m^2$ corresponds to roughly $5.6 \times 10^5$ molecules, with ~ 3000 molecules at the boundary. A PnQ number fraction of 0.013 then would correspond to a grain surrounded by approximate 2 layers of PnQ, as suggested in Fig. 4b, that would effectively no longer transport holes. The real system of course does not reach zero mobility with increasing PnQ, as shown in Fig. 1b. This may be explained by the result illustrated in Fig. 2a and Fig. 2b, that PnQ and Pn effectively separate for high concentrations of PnQ, precluding the complete loss of mobility in either of the limiting-case models discussed. This study cannot differentiate between these two simplistic models but both indicate the relatively large effects of relatively small concentrations of PnQ.

## Conclusions

This data reveals three important facts. First, Pn and its primary impurity, PnQ, have phase separate above a PnQ number density ~0.008. The phase-separated materials have very different growth modes (layer-by-layer mode for Pn, *vs.* 3-d growth for PnQ). However, the decrease in Pn mobility with PnQ crystallite volume is more rapid than linear, indicating that chemical segregation cannot be complete at low impurity concentrations. The second fact is that for samples with a small PnQ impurity level, no more than a number fraction of 0.008, Pn thin film growth habits are not measurably affected and similar ultra-thin film and bulk film morphologies are observed for these films. However, strong decreases in mobility are observed in this range, indicating that direct effects of PnQ, rather than changes in grain boundary density, are limiting the mobility. These chemical effects could be in the form of charge traps due to: (i) local potential changes due to individual structural imperfections created by PnQ molecules in the Pn crystalline phase, or (ii) a concentration of PnQ molecules at the natural grain boundaries of Pn thin film structure. The first hypothesis is supported by the recent observation by EFM of charge traps inhomogeneously distributed in Pn films (and not only confined to grain boundaries)[32, 34], and by the observation of strong mobility reductions due to impurities in single crystalline Pn[4]. Finally, strong perturbation of the Pn growth habit is observed above an impurity number fraction of 0.008. In this range, the continuing strong decreases in mobility reflect some combination of effects of the degraded morphology and impurity scattering.

In summary, we correlated the dependencies of the growth morphology and the field-effect mobility of Pn OTFTs on the percentage of added impurity PnQ present in the source material. The results show that PnQ impurities degrade device performance well before affecting Pn crystal growth habit. Thus improved growth quality alone cannot be used as a predictor of improved device performance.

ACKNOWLEDGMENT

This work was supported by the UMD NSF-MRSEC under grant DMR 05-20471, by the LPS, and by NIST under contract #70NANB6H6138 with research infrastructure supported by the UMD-CNAM. The NSF-MRSEC SEF was used in obtaining the data presented. We thank D.R. Hines, M. Breban and V. Ballarotto for help in developing preparation and characterization procedures, and to T.L. Einstein, S. Sundaresan and A.E. Southard for helpful discussions.

FIGURE CAPTIONS

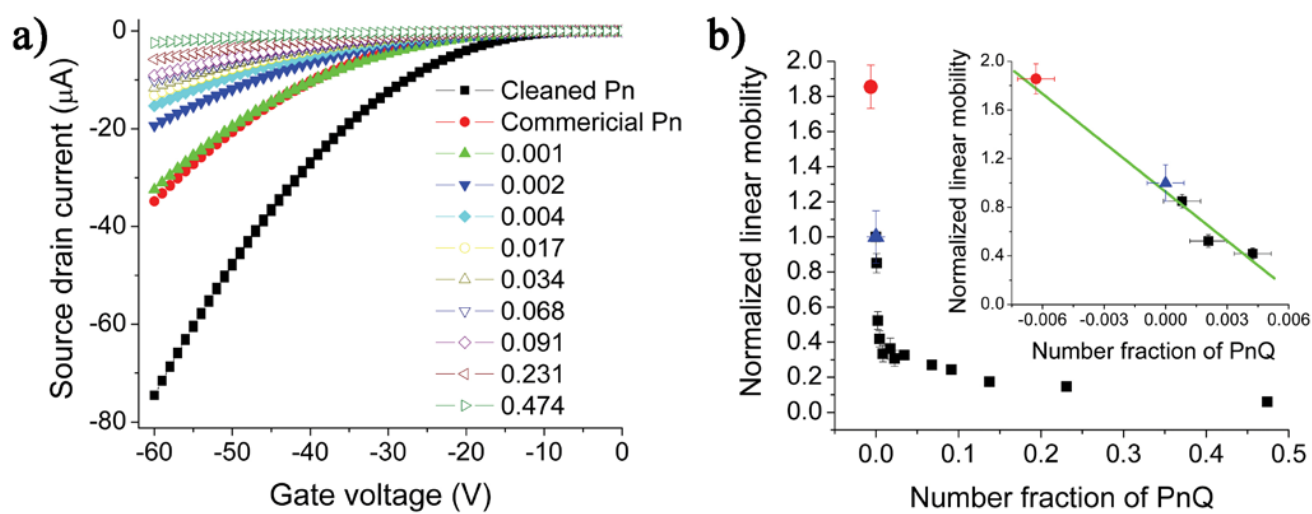

**Figure 1**: Color online (a) Averaged transfer curves of 50 nm film OTFTs prepared: cleaned Pn, commercial Pn, and PnQ/Pn admixtures with a number fraction of added PnQ ranging from 0.000 to 0.474. The gate voltage was swept at a constant $V_{s-d}$ = - 40V. (b) Normalized linear mobility of cleaned Pn (●, full circle), commercial Pn (▲, full triangle) and PnQ/Pn admixtures from 0.006 to 0.474 (■, full squares) *versus* number fraction of added PnQ. The inset is a zoom-in of the lower PnQ concentrations. The solid line is a linear fit to the PnQ/Pn admixture device data.

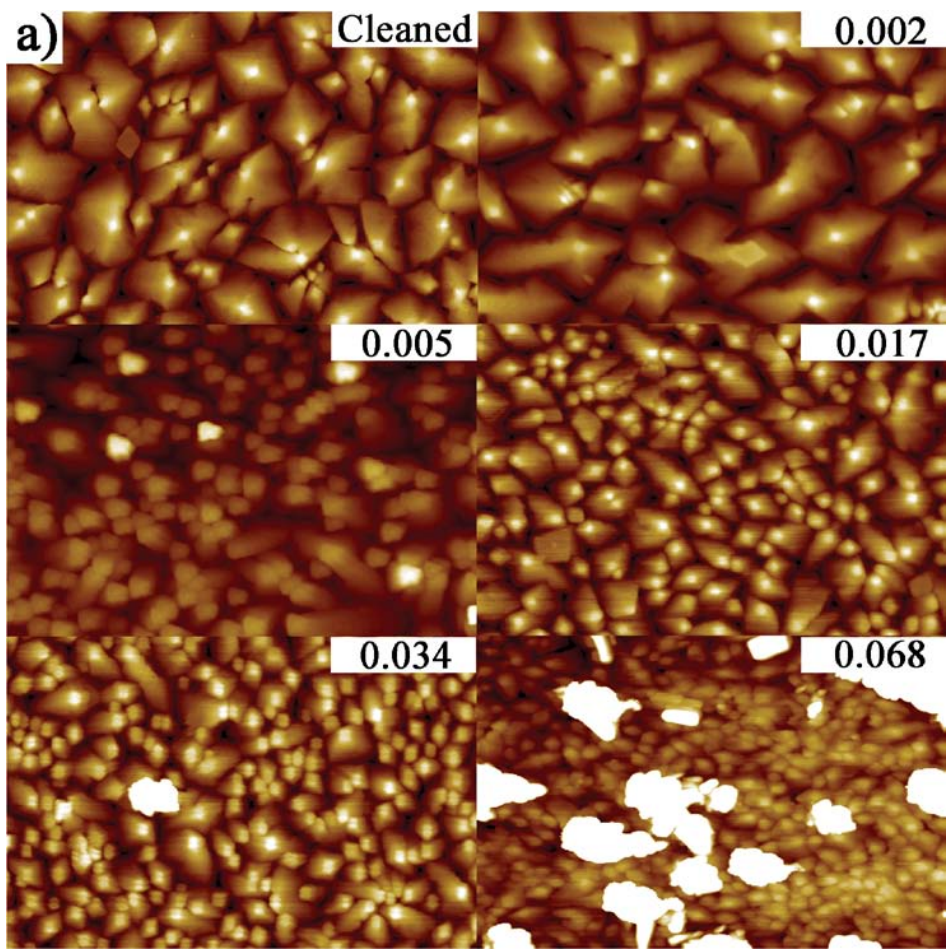

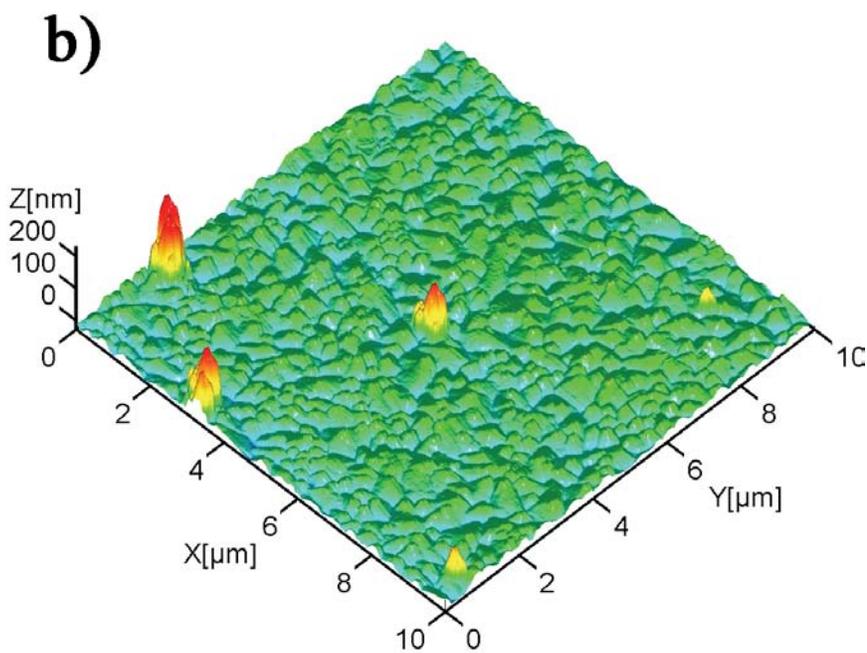

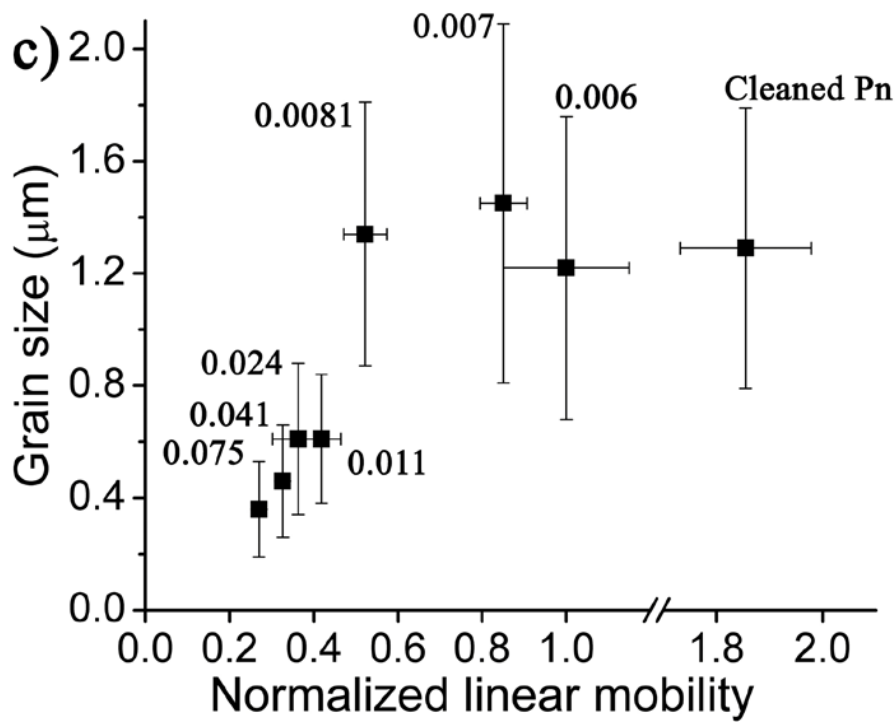

Figure 2: Color online(a) 7.5 μm by 5.0 μm AFM images of nominal 50 nm thick films prepared using cleaned Pn, commercial Pn (or +0.0%PnQ) and PnQ/Pn admixtures with a percentage of added PnQ from +0.1 to +7.5% (net number fraction from 0 to 0.068). (b) 3D image of a nominal 50nm thick film using 0.034 PnQ number fraction source material (c) Grain size as a function of normalized linear mobility for films presented in Fig. 2a. Labels on each data point indicate the PnQ number fraction.

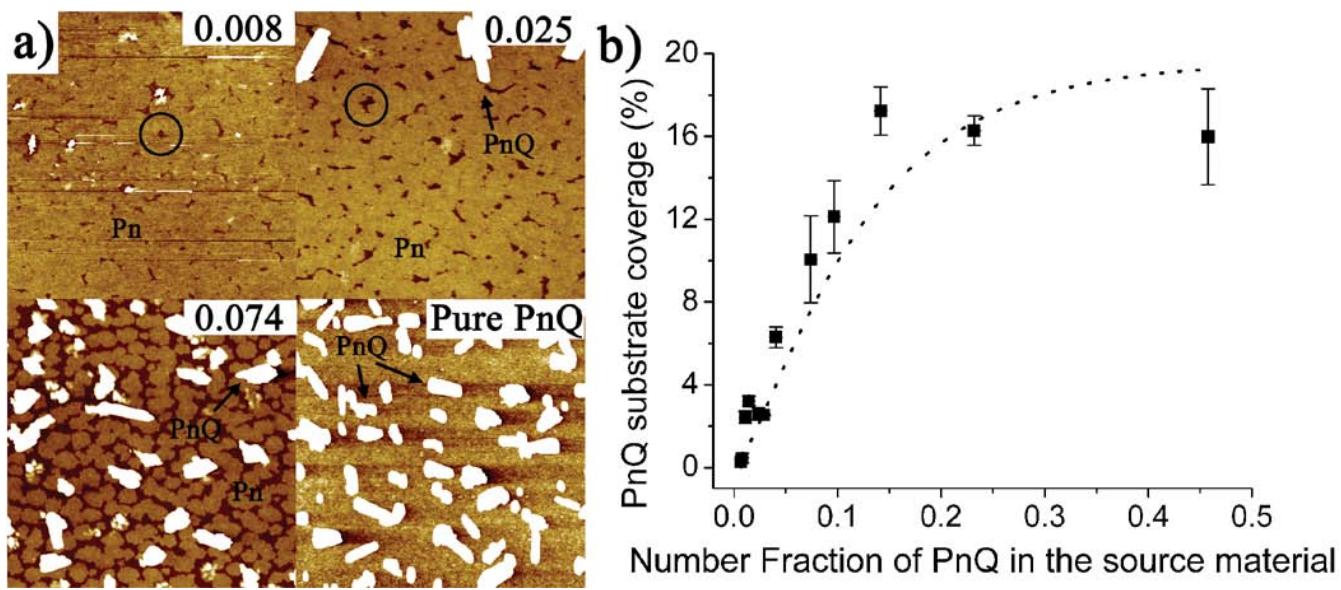

**Figure 3:** Color online (a) AFM images of films with a deposition time equivalent to one Pn monolayer, prepared using as a source material PnQ/Pn admixture with a number fraction of PnQ of 0.008, 0.025, 0.074 and with pure PnQ. Regions showing bare $SiO_2$ are circled. (b) Fraction of the substrate covered with PnQ crystallites as a function of number fraction of PnQ. The broken black line is a guide to the eye.

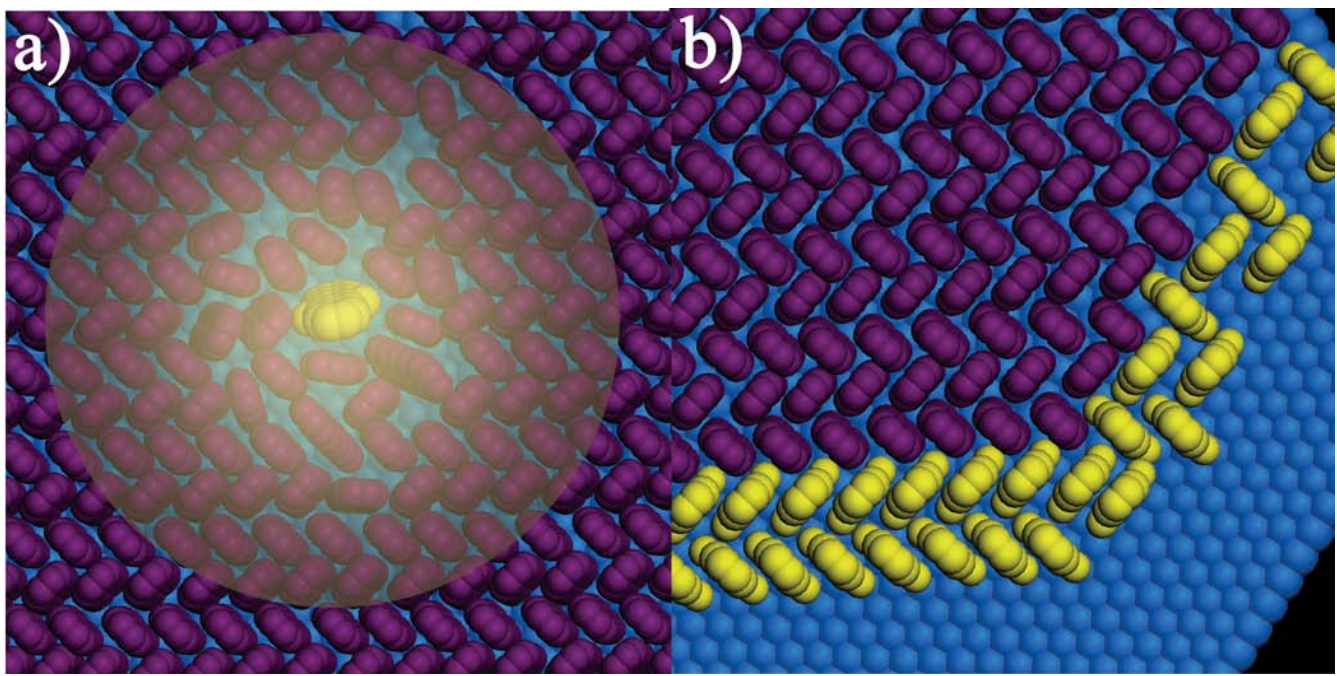

**Figure 4:** Color online (a) An illustration of a monolayer of Pn with a single PnQ molecule disrupting the lattice. (b) An illustration of a Pn grain boundary surrounded by 2 layers of PnQ.

TABLES.

**Table 1.** Nucleation density as a function of impurity concentration in source material

| Absolute PnQ in Source (%) | Number Fraction of PNQ | Nucleation Site Density $(1/\mu m)^2$ |
|---|---|---|
| 0.0 ± 0.1 | 0.000 ± 0.001 | 3.9 ± 0.2 |
| 0.7 ± 0.1 | 0.006 ± 0.001 | 3.8 ± 0.1 |
| 0.8 ± 0.1 | 0.008 ± 0.001 | 3.1 ± 0.3 |
| 1.2 ± 0.1 | 0.011 ± 0.001 | 9.7 ± 0.9 |
| 2.0 ± 0.1 | 0.018 ± 0.001 | 5.1 ± 1.5 |
| 3.2 ± 0.1 | 0.029 ± 0.001 | 6.1 ± 1.7 |
| 4.5 ± 0.1 | 0.041 ± 0.001 | 5.0 ± 0.8 |